\documentclass[twocolumn]{aastex62}

\usepackage{graphicx}
\usepackage{natbib}
\usepackage{amsmath}
\usepackage{url}
\shorttitle{Sub-Saturns and Gas Giants}
\shortauthors{Lee}

\begin{document}

\title{The Boundary Between Gas-rich and Gas-poor Planets}

\correspondingauthor{Eve J. Lee}
\email{evelee@caltech.edu}

\author{Eve J. Lee}
\affil{TAPIR, Walter Burke Institute for Theoretical Physics, Mailcode 350-17, Caltech, Pasadena, CA 91125, USA}

\begin{abstract}
Sub-Saturns straddle the boundary between gas-rich Jupiters and gas-poor super-Earths/sub-Neptunes. Their large radii (4--8$R_\oplus$) suggest that their gas-to-core mass ratios range $\sim$0.1--1.0.
With their envelopes being as massive as their cores, sub-Saturns are just on the verge of runaway gas accretion; they are expected to be 
significantly less populous than gas giants.
Yet, the observed occurrence rates of sub-Saturns and Jupiters are comparable within $\sim$100 days. 
We show that in these inner regions of planetary systems, 
the growth of 
sub-Saturns/Jupiters 
is ultimately limited by 
local and global 
hydrodynamic flows---runaway accretion terminates and the formation of gas giants is suppressed. 
We derive a simple analytic formula for the 
local
hydrodynamic accretion rate---an expression that has been previously reported only as an empirical fit to numerical simulations. Evolving simultaneously the background disk gas and the gas accretion onto planetary cores, we find that both the ubiquity of super-Earths/sub-Neptunes and the rarity of gas-rich planets are best explained if an underlying core-mass distribution is peaked at $\sim$4.3$M_\oplus$. 
Within a finite disk lifetime $\sim$10 Myrs, massive cores ($\gtrsim 10M_\oplus$) can become either gas-poor or gas-rich depending on when they assemble, but smaller cores ($\lesssim 10M_\oplus$) can only become gas-poor. This wider range of possible outcomes afforded by more massive cores may explain why metal-rich stars harbor a more diverse set of planets.
\end{abstract}

\section{Introduction} \label{sec:intro}
The theory of core accretion provides one of the most successful explanations of how planets acquire their gaseous envelopes \citep[e.g.,][]{Perri74,Mizuno80,Stevenson82}.
Time-dependent accretion models such as \citet{Pollack96} describe the birth of planets in three phases. First, rocky cores are amassed from the solid disk (phase 1). They accrete their gaseous envelopes at a rate regulated by internal cooling (phase 2), and blow up into gas giants as the gas accretion rate ``runs away'' in response to the atmosphere's self-gravity (phase 3). The planet becomes a gas giant only if phase 2 ends and phase 3 begins within the lifetime of the gas disk. The duration of phase 2 is largely governed by the mass of the core, with more massive cores triggering runaway faster.

The standard picture of core accretion expects a bimodal population of planets: gas-poor, predominantly rocky planets vs.~gas giants \citep[see, e.g.,][]{Ida04}. Even modern population synthesis models such as \citet{Mordasini18} report a pronounced peak in the population of gas giants (see, e.g., their Figure 10).\footnote{While \citet{Mordasini18} report an absence of peak of Jupiter-sized planets within 0.27 AU, it is likely that the peak will reappear when the sample is extended to $\sim$1 AU (see, e.g., their Figure 9). Modern estimates of planet occurrence rates from {\it Kepler} data that extend to $\sim$300 days still report a lack of any peak at Jupiter radii \citep[e.g.,][]{Petigura18}.} 
Observations of {\it Kepler} planets challenge these standard 
expectations. 
The distribution of planetary radii is flat beyond $\sim$4$R_\oplus$ \citep{Petigura13}. We see just as many sub-Saturns on the verge of runaway (4--8$R_\oplus$, envelope mass fractions of $\sim$10--100\%) as we do gas giants \citep{Dong13,Petigura18}. 

A related question is what limits the growth of gas giants. Once the planet has begun its runaway growth, its cooling timescale shortens catastrophically. Eventually, hydrodynamics, rather than thermodynamics dictate
how fast the gas envelopes grow \citep[see, e.g.,][for qualitative descriptions]{Bodenheimer00,Marley07}. In an infinite medium of gas, the maximum rate of gas accretion is set by the classical Bondi accretion \citep{Bondi52}. In protoplanetary disks, gas flows around cores become complicated by three-body dynamics and the geometry of the disk. Flow velocities can be set by the shear velocity at the accretion surface rather than the local sound speed (unlike in star-forming molecular clouds, turbulence in protoplanetary disks is expected to be subsonic; \citealt{Pinte16,Flaherty17}). 
Once the Hill radius of the planet exceeds the local disk scale height---i.e., exceeding the ``thermal'' mass---the gravitational perturbation the planet exerts at Lindblad resonances can become non-linear, significantly altering the structure of the disk gas \citep{lin93}.
\citet{Tanigawa07} and \citet{Tanigawa16} computed the final mass of gas giants by tracking gas accretion onto cores, 
accounting for the gas cooling, hydrodynamic flows, and the global disk accretion by viscous diffusion.
They reported the final mass over a range of disk parameters such as the Shakura-Sunyaev viscous parameter $\alpha$ \citep{Shakura73} and the initial disk gas surface density, predicting more massive planets in disks with high $\alpha$ and high gas densities (i.e., large disk accretion rates).

It is possible that the observed diversity in planetary population is a reflection of the distribution of disk gas parameters. However, how quickly the planetary cores can trigger the runaway growth is most sensitively determined by their masses. 
In this paper, we search for the underlying distribution of core mass that yields both the ubiquity of sub-Neptunes and the similar occurrence rates of sub-Saturns and gas giants.  
We assess the effect of hydrodynamic flows in shaping the population of gas-rich planets.
This paper is organized as follows.
We start with a review of various modes of gas accretion in Section \ref{sec:calc}, providing an analytic, order-of-magnitude calculations of accretion rates set by cooling, as well as by hydrodynamic flows. Section \ref{sec:distrb} describes the method we use to compute the inferred and the model cumulative distribution function (CDF) of gas-to-core mass ratios (GCR) using the observed planet occurrence rates and the model of envelope growth, respectively. The best-fit core-mass distribution that brings the observed and the model CDF of GCR into agreement is derived in Section \ref{sec:best-fit}. The physics that governs the maximum gas mass of a given core is described in Section \ref{sec:outcome}. We summarize in Section \ref{sec:summary}, highlighting the implications of our results and avenues for improvement.

\section{Model of Envelope Growth} \label{sec:calc}

\subsection{Cooling-limited Gas Accretion}
\label{ssec:cooling}

Cores accrete as much gas as can cool.
In particular, the timescale of accretion is set by the cooling timescale of the inner convective zone of the envelope. Using basic principles of thermodynamics, \citet{Lee15} derived an analytic scaling relationship between the gas-to-core mass ratio, the accretion time, the atmospheric metallicity, and the core mass (see also \citealt{Ginzburg16}). We briefly summarize below the key physical ingredients. The timescale of accretion is simply
\begin{equation}
    t \sim t_{\rm cool} \sim \frac{|E|}{L},
\end{equation}
where $E$ is the total energy of the envelope and $L$ is the cooling luminosity. From hydrostatic equilibrium, $E$ is, to order unity, given by the gravitational potential energy of the gaseous envelope bound to the central core:
\begin{equation}
    E \sim \frac{GM_{\rm core}M_{\rm gas}}{R_{\rm core}},
\end{equation}
where $G$ is the gravitational constant, $M_{\rm core}$ is the core mass, $M_{\rm gas}$ is the gas mass, and $R_{\rm core}$ is the radius of the core. The important length scale here is $R_{\rm core}$, because envelope masses are centrally concentrated \citep{Lee14}. This central concentration follows from the development of a steep adiabat within the inner convective zone. There, the temperature rises beyond $\sim$2500 K, hot enough to dissociate hydrogen molecules. Energy is spent in dissociating hydrogen molecules instead of heating up the gas, effectively driving the temperature gradient close to zero and steepening the density profile.

The cooling luminosity $L$ is given by the radiative diffusion at the radiative-convective boundary:
\begin{equation}
    L = \frac{64 \pi G (M_{\rm core} + M_{\rm gas})\sigma T_{\rm rcb}^3 \mu_{\rm rcb} m_{\rm H} \nabla_{\rm ad}}{3k\rho_{\rm rcb}\kappa_{\rm rcb}},
\end{equation}
where $\sigma$ is the Stefan-Boltzmann's constant, $T_{\rm rcb}$ is the temperature, $\mu_{\rm rcb}$ is the mean molecular weight, $m_{\rm H}$ is the mass of a hydrogen atom, $\nabla_{\rm ad}$ is the adiabatic gradient, $\rho_{\rm rcb}$ is the density, and $\kappa_{\rm rcb}$ is the opacity, all evaluated at the radiative-convective boundary (rcb). The rate of cooling is set by the properties of the rcb because this boundary acts as a thermal bottleneck. The upper radiative layer functions as a lid that regulates the rate at which energy escapes out of the inner convective zone. Radiative cooling is set by the local temperature gradient, which, by definition, is maximized at the rcb---the properties of the rcb set the maximum rate of energy transport.

For dusty envelopes, the rcb emerges at the hydrogen molecule dissociation front. Free hydrogen atoms combine with free electrons to form H- ions. The opacity due to the bound-free transition of H- ions rises steeply with temperature, ensuring convection prevails in the deeper envelopes. This effectively fixes $T_{\rm rcb}$ to 2500 K. From tabulated opacities that consider the formation, the dissociation, and the chemical reaction of different grains and gas molecular species, we obtain $\kappa(H-) \propto \rho^{0.5} T^{7.5}$ \citep{Ferguson05}. We can now write an analytic scaling relationship between the gas mass, the time, and the core mass:
\begin{equation}
    \frac{M_{\rm gas}}{M_{\rm core}} = 0.09 \left(\frac{\Sigma_{\rm neb}}{13\,{\rm g\,cm^{-2}}}\right)^{0.12} \left(\frac{M_{\rm core}}{20\,M_\oplus}\right)^{1.7} \left(\frac{t}{0.1\,{\rm Myrs}}\right)^{0.4}
    \label{eq:gcr}
\end{equation}
where $\Sigma_{\rm neb}$ is the nebular gas surface density and 0.09$(\Sigma_{\rm neb}/13\,{\rm g\,cm^{-2}})^{0.12}$ is the normalization factor from numerical calculations \citep{Lee14,Lee16}. Our calculation assumes dust grains contribute to the mean opacity in the upper layers of the planetary atmosphere, and the dust grain size distribution follows that of the interstellar medium. While dust grains in the atmosphere may coagulate and rain out, the advection of the surrounding gas can bring fresh supplies of dust to the upper layers of the bound envelope. Detailed three-dimensional hydrodynamic calculations report a three-layer structure: the innermost convective zone, radiative layer, and the uppermost advective zone \citep[e.g.,][]{Lambrechts17}. These studies report the advection zone reaches down to about a third of the Bondi radius ($R_{\rm Bondi}$) or Hill radius ($R_{\rm Hill}$). In deriving equation \ref{eq:gcr}, we have modified the numerical calculation of \citet{Lee14} to set the outermost radius to ${\rm min}(R_{\rm Hill}, R_{\rm Bondi})/3$ which decreases the final gas-to-core mass ratio by about a factor of 3. 

We have also assumed the atmospheric metallicity to be solar ($Z=0.02$). Up to $Z=0.2$, gas accretion rates drop with increasing metallicity as metallic species make the envelopes more opaque and delay cooling. Beyond $Z=0.2$, envelopes become so heavy that their gravitational contraction outweighs the enhancement in opacity. We do not consider the effect of metallicity here as ``atmopsheric'' metallicity is poorly constrained. 
Measurements of transit spectroscopy have been obtained for only a handful of planets (e.g., GJ 436 b, \citealt{Knutson14}; GJ 1214 b, \citealt{Kreidberg14}; HAT-P-11b, \citealt{Fraine14}; HAT-P-26b, \citealt{Wakeford17}). Although these observations suggest that planetary envelopes are significantly enhanced in metallic content, it is not clear whether such enhancement is uniform throughout the envelope or if it varies with depth. Gradients in the abundance of heavy elements have been shown to alter the thermal structure and evolution of gas giants \citep[e.g.,][]{Leconte12,Vazan16} and sub-Neptunes \citep[e.g.,][]{Bodenheimer18}, but it is not clear whether such gradients (and any metallic enhancement) are established during or after the initial build up of planetary envelopes.
Far from the host star where disks are cold enough for ice to condense, high-metallicity envelopes can significantly hasten the growth of the envelope \citep[e.g.,][]{Venturini15}. We do not consider most super-Earths to have initially formed as full-fledged planets farther out then migrated in as large-scale migration of planetary bodies have trouble explaining (1) the lack of pile-up closest to the star \citep{Lee17}; (2) the majority of Kepler planets being significantly outside of resonance \citep{Fabrycky14,Deck15}; and (3) the high bulk densities of bare rocky planets being inconsistent with icy bodies \citep{Owen17}. Some planetary systems, such as TRAPPIST-1 \citep{Gillon17}, betray signatures of migration with their complex web of resonances, but they are likely a minority. 

\begin{figure}
    \centering
    \includegraphics[width=\columnwidth]{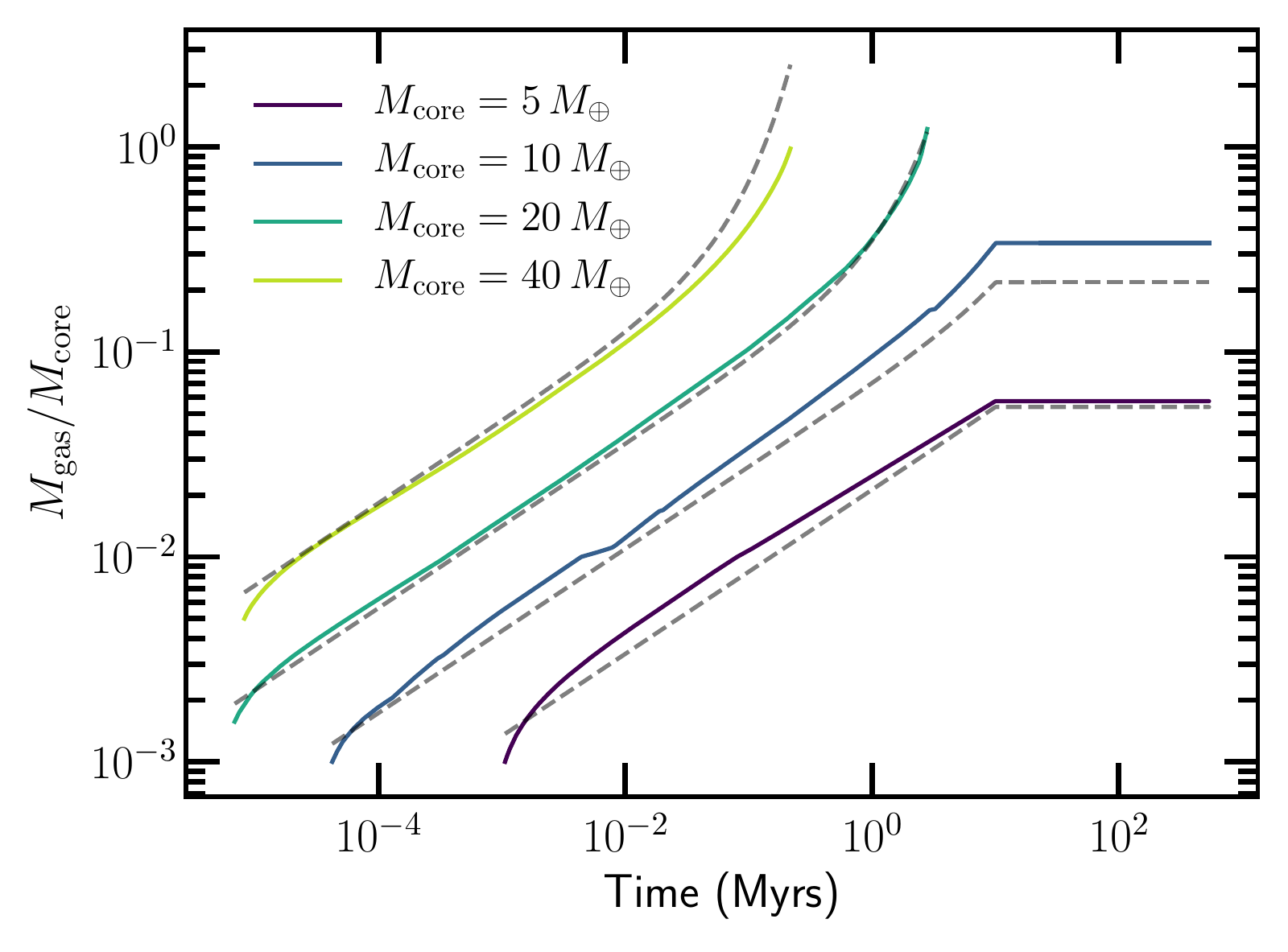}
    \caption{Comparison between equation \ref{eq:gcr_exp} (dashed lines) and numerical results (solid lines). Numerical calculations are performed at 0.1 AU with $\Sigma_{\rm neb} = 39.4\,{\rm g}\,{\rm cm}^{-2}$, $T_{\rm disk} = 1000\,{\rm K}$, and dusty envelopes. The outer boundary of the envelope is taken at 0.3$\times {\rm min}(R_{\rm Hill}, R_{\rm Bondi})$. We truncate any curve that extends beyond 10 Myr, the typical timescale over which the disk gas disperses. Overall, the analytic formula agrees with the numerical result within factors of order unity.}
    \label{fig:gcr_comp}
\end{figure}

Once the gas mass becomes comparable to the core mass, the self-gravity of the envelope shortens the cooling timescale at a catastrophic rate and runaway accretion ensues. We approximate this runaway growth as an exponential:
\begin{align}
    \frac{M_{\rm gas}}{M_{\rm core}} &= 0.09 
    \left(\frac{\Sigma_{\rm neb}}{13\,{\rm g\,cm^{-2}}}\right)^{0.12}
\left(\frac{M_{\rm core}}{20\,M_\oplus}\right)^{1.7} \left(\frac{t}{0.1\,{\rm Myrs}}\right)^{0.4} \nonumber \\
    &\times \exp\left[\frac{t}{2.2\,{\rm Myrs}}\left(\frac{M_{\rm core}}{20\,M_\oplus}\right)^{4.2}\right].
    \label{eq:gcr_exp}
\end{align}
where 2.2 Myr $(M_{\rm core}/20\,M_\oplus)^{-4.2}$ is the core-mass-dependent runaway timescale. Figure \ref{fig:gcr_comp} demonstrates how our exponential approximation agrees with the numerical calculation within factors of order unity.
In the limit of $M_{\rm gas} \gg M_{\rm core}$, cooling timescales shorten with heavier cores, leading to super-exponential growth of gas mass \citep{Ginzburg19}. In close-in orbits, we are safe with our assumption of exponential growth since runaway growth is almost always stopped before $M_{\rm gas} \gtrsim 10\,M_{\rm core}$.

Our discussion of accretion by cooling assumes spherically symmetric flow, most appropriate for planets whose bound radius is within the local disk scale height. Once a planet enters the runaway regime and exceeds the thermal mass (i.e., when its Hill sphere exceeds the local disk scale height), it can significantly perturb the ambient disk gas and the flow geometry likely becomes highly aspherical. It may even be that the global disk gas accretion lags behind local gas accretion onto the planetary cores. We discuss these considerations in the next section.

\subsection{Hydrodynamic Considerations}
\label{ssec:hydro}

Numerical calculations that consider the growth of gas giants post runaway report accretion rates empirically fit to simulation results \citep[e.g.,][]{Tanigawa02,Lissauer09,Dangelo13}.
\citet{Tanigawa16} provide a best-fit scaling relationship for planets whose $R_{\rm Hill} \gtrsim H$ (i.e., super-thermal mass) where $H = c_{\rm s}/\Omega$ is the gas disk scale height, $c_{\rm s}$ is the sound speed, and $\Omega$ is the Keplerian orbital frequency. We sketch below an order-of-magnitude estimate of how such scaling relationship may be obtained. 

Mass accretion rate can be expressed as $\dot{M} \sim \rho_{\rm in} A v_{\rm in}$ where $\rho_{\rm in}$ and $v_{\rm in}$ are the density and the velocity of the incoming flow, respectively, and $A$ is the cross section at which the flow contacts the planet. Since $R_{\rm Hill} \gtrsim H$, $A \sim 2\pi R_{\rm Hill} H$. 
The shear velocity at the Hill radius dominates the background sound speed so the flows shock near the Hill sphere, dissipate energy and fall onto the planet. 
At the Hill radius, the free-fall velocity of the gas is $v_{\rm in} \sim \Omega R_{\rm Hill}$.
Assuming the shock to be isothermal (as was assumed in the simulations of \citealt{Tanigawa02}), we take the post-shock density $\rho_{\rm in} \sim \rho_{\rm neb} (\Omega R_{\rm Hill}/c_{\rm s})^2$ where $c_{\rm s}$ is the sound speed and $\rho_{\rm neb}$ is the background nebular density. 
The expected mass accretion rate is then
\begin{align}
    \dot{M}_{\rm hydro} &\sim 2\pi R^2_{\rm Hill} H \rho_{\rm neb} \left(\frac{\Omega R_{\rm Hill}}{c_{\rm s}}\right)^2 \Omega \nonumber \\
    &\propto \left(\frac{M_{\rm p}}{M_\star}\right)^{4/3} \Sigma_{\rm neb} \left(\frac{a}{H}\right)^2 a^2 \Omega,
    \label{eq:Mdot_hydro}
\end{align}
where $M_{\rm p} \equiv M_{\rm gas} + M_{\rm core}$ is the total mass of the planet and $a$ is the orbital distance. This is in agreement with equations 7 and 8 of \citet{Tanigawa16}, which we rewrite as
\begin{equation}
\dot{M}_{\rm hydro} = 0.29 \left(\frac{M_{\rm p}}{M_\star}\right)^{4/3} \Sigma_{\rm neb} \left(\frac{a}{H}\right)^2 a^2 \Omega.
\label{eq:Mdot_hydro_norm}
\end{equation}

Our approximation $A \sim 2\pi R_{\rm Hill} H$ assumes accretion in the equatorial region. While such an approximation is applicable for a two-dimensional calculation as \citet{Tanigawa02} have performed, three-dimensional simulations generally find accretion along the pole and decretion along the equator \citep{Tanigawa12_3D,Dangelo13,Cimerman17}.\footnote{Isothermal three-dimensional (3D) simulations report no bound atmosphere as gas flows cycle into and out of the Hill/Bondi sphere \citep[e.g.,][]{Fung15,Ormel15}. Relaxing the assumption of isothermality largely suppresses the degree of recycling \citep[e.g.,][]{Dangelo13,Cimerman17,Lambrechts17}. Planetary cores amass their envelopes by cooling the gas so that by definition, the interior of planetary envelopes would be at lower entropy than the outer nebula. It follows that the flows from the disk gas will be buoyed away, unable to penetrate the deeper layers of the envelopes \citep{Kurokawa18}.} 
Numerical simulations that study in detail the formation of circumplanetary disks also report at minimum $\sim$90\% of the gas accretion onto planetary cores is in the polar direction \citep[e.g.,][]{Szulagyi14}.
It may be more appropriate to take $A$ as some fraction of $4\pi R_{\rm Hill}^2$ where the fraction needs to be determined by detailed hydrodynamic calculations for super-thermal mass planets.

It should also be noted that the shock may be adiabatic, in which case, 
$\rho_{\rm in} \sim \rho_{\rm neb} (\gamma + 1)/(\gamma - 1 + 2/{\cal M}^2)$ where $\gamma$ is the adiabatic index of the nebular gas and 
${\cal M} \sim \Omega R_{\rm Hill}/c_{\rm s}$
is the shock Mach number.
For the shock to be isothermal, we need the gas to radiate away the kinetic energy of the infalling gas within the freefall time; at the Hill radius, the freefall time is simply the local orbital time $\sim$ $\Omega^{-1}$. Assuming the surface of the planet cools as a blackbody, we can write the shock cooling time as $t_{\rm cool,shock} \sim \Sigma_{\rm neb} A v_{\rm in}^2 / A \sigma T^4$ where $\Sigma_{\rm neb}$ is the local gas surface density, $A$ is the shock cross section, $v_{\rm in} = \Omega R_{\rm Hill}$, $\sigma$ is the Stefan-Boltzmann constant, and $T$ is the surface temperature of the planet, taken as the midplane temperature of the disk. We evaluate 
\begin{align}
t_{\rm cool,shock}\Omega &\sim \Sigma_{\rm neb} \Omega^3 R_{\rm Hill}^2 / \sigma T^4 \nonumber \\
&\sim 0.04 \left(\frac{0.1\,{\rm AU}}{a}\right)^{5/2} \left(\frac{M_\star}{M_\odot}\right)^{5/6} \left(\frac{M_{\rm p}}{5\,M_\oplus}\right)^{2/3} \nonumber \\ 
&\times \left(\frac{\Sigma_{\rm neb}}{13\,{\rm g\,cm^{-2}}}\right) \left(\frac{1000\,{\rm K}}{T}\right)^{4}.
\end{align}
For gas-poor nebula, the approximation of isothermal shock is valid but for gas-rich nebula, close to the star, adiabatic approximation may be more appropriate. We expect a different scaling relationship for $\dot{M}_{\rm hydro}$ if the shock is adiabatic: $\dot{M}_{\rm hydro} \propto R_{\rm Hill}^2 H \rho_{\rm neb} \Omega \propto (M_{\rm p}/M_\star)^{2/3} \Sigma_{\rm neb} a^2 \Omega$. Future numerical simulations that consider non-isothermal gas accretion onto massive cores would be welcome to verify our calculations and to constrain the normalization. In the absence of such 
calculation,
we assume isothermal shock throughout the paper.

The nebular density in equation \ref{eq:Mdot_hydro} is evaluated at $R_{\rm Hill}$ of the planet. Super-thermal mass planets are expected to perturb the surrounding nebula and open up 
deep gaps.
Using the empirically determined gap depth derived by \citet{Duffell13} and \citet{Fung14}, \citet{Tanigawa16} evaluate $\Sigma_{\rm neb} = \Sigma_{\rm bg}/(1+0.034K)$ with
\begin{equation}
    K = \left(\frac{H}{a}\right)^{-5}\left(\frac{M_{\rm p}}{M_\star}\right)^2\alpha^{-1}
    \label{eq:K}
\end{equation}
where $\alpha$ is the Shakura-Sunyaev viscous parameter, and $\Sigma_{\rm bg}$ is the unperturbed background disk gas surface density. The depletion factor $K$ can be derived analytically by equating the one-sided Lindblad torque of a planet pushing on the gas to the viscous torque of the disk refilling the gas \citep[see][their Section 4.3]{Fung14}.\footnote{For planets with $R_{\rm Hill} > H$, it is not clear whether the gap extends all the way to $R_{\rm Hill}$. While a more careful investigation of the gap profile in the super-thermal regime is warranted, we verify that in the most pessimistic limit of $K=0$, the final envelope mass fractions change only by order unity and only for very massive cores ($M_{\rm core} > 40\,M_\oplus$).}

\begin{figure}
    \centering
    \includegraphics[width=\columnwidth]{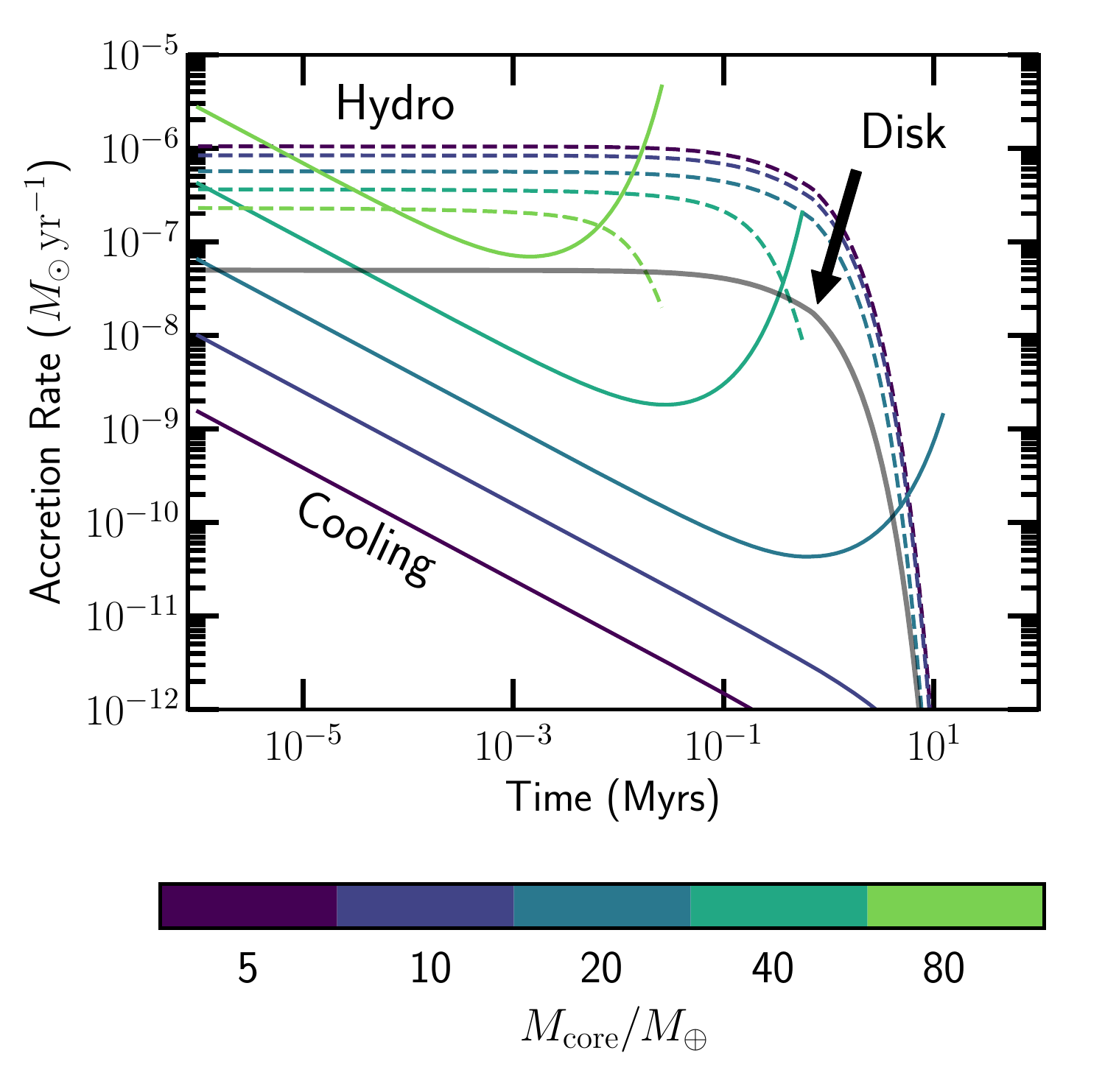}
    \caption{Comparison of gas accretion rates between accretion by cooling ($\dot{M}_{\rm cool}$, solid color curves; equation \ref{eq:Mdot_cool}), hydrodynamic flows ($\dot{M}_{\rm hydro}$, dashed lines; equations \ref{eq:Mdot_hydro_norm} and \ref{eq:K}), and global disk accretion ($\dot{M}_{\rm disk}$, solid gray curve; equation \ref{eq:Mdisk_dot}). Both hydrodynamic flows and disk accretion assume $\alpha=10^{-3}$. For $M_{\rm core} < 20 M_\oplus$, $\dot{M}_{\rm hydro}$ declines sharply beyond $\sim$1 Myr, just like $\dot{M}_{\rm disk}$. This drop reflects the rapid dispersal of inner disk gas, potentially by photoevaporation cutting off the gas
    inflow from beyond $\sim$1 AU \citep{Owen11}.
    For high mass cores, $\dot{M}_{\rm hydro}$ drops at earlier times because they trigger the runaway earlier and carve out a deep gap in the disk, reducing significantly the local gas surface density. Low-mass ($\lesssim 20M_\oplus$) cores build their gaseous envelopes entirely by gas cooling. Accretion onto high-mass cores is initially limited by the global disk gas accretion. Once the planet gains enough mass to carve out a deep gap in the disk, their growth is limited by hydrodynamic flows.}
    \label{fig:Mdot_comp}
\end{figure}

When the gas accretion onto a core is limited by hydrodynamic flows, the rate of accretion becomes sensitive to the background gas density. We assume steady-state accretion disk and use the best-fit parameters from \citet{Owen11}, fitted to the observed accretion rates of T Tauri stars as a function of age:
\begin{equation}
    \frac{\Sigma_{\rm bg}(a, t)}{\rm g\, cm^{-2}} = 
    \begin{cases}
        3\times 10^{4}\left(\frac{a}{0.2\,{\rm AU}}\right)^{\frac{-15}{14}}\left(\frac{t}{t_{\rm visc}}+1\right)^{\frac{-3}{2}}, & t < t_{\rm visc} \\
        10^{4}\left(\frac{a}{0.2\,{\rm AU}}\right)^{\frac{-15}{14}}\exp\left(-\frac{t}{t_{\rm visc}}\right), & t > t_{\rm visc}.
    \end{cases}
    \label{eq:Sigma_gas}
\end{equation}
where we take $t_{\rm visc}=0.7$ Myr and the temperature profile of 1000\,K\,(a/0.1\,AU)$^{-3/7}$. The first branch $t < t_{\rm visc}$ corresponds to a steady-state, viscously spreading disk \citep{lynden_bell74,hartmann98}.
After $t \sim t_{\rm visc}$, mass loss by photoevaporative wind dominates the disk evolution, carves out a gap at $\sim$1 AU and decouples the inner disk from the outer disk. The inner disk disperses rapidly on a viscous timescale evaluated at the gap radius $\sim$1 AU \citep{Owen11}.

Assuming $\alpha=10^{-3}$, we estimate the global disk gas accretion rate as
\begin{align}
    \dot{M}_{\rm disk} &= 3\pi\alpha c_{\rm s}H\Sigma_{\rm bg}(a, t) \nonumber \\
    &\sim
    \begin{cases}
    4\times 10^{-8}\,M_\odot\,{\rm yrs}^{-1}\left(\frac{t}{t_{\rm visc}}+1\right)^{-3/2}, & t < t_{\rm visc} \\
    10^{-8}\,M_\odot\,{\rm yrs}^{-1}\exp\left(-\frac{t}{t_{\rm visc}}\right), & t > t_{\rm visc}.
    \end{cases}
    \label{eq:Mdisk_dot}
\end{align}
We note that both 
$\alpha$ and the normalization of $\Sigma_{\rm bg}$ are chosen to match the observed accretion rates onto T Tauri stars \citep[see, e.g.,][their Figure 5]{Owen11}.

Figure \ref{fig:Mdot_comp} compares gas accretion rates from cooling, hydrodynamic flows, and the global disk accretion. Cores less massive than 
$\sim$10$M_\oplus$
always accrete gas by cooling. For more massive cores, gas delivery is mostly limited by the global disk accretion. 
Local hydrodynamic 
flows only become important once the planet triggers runaway gas accretion, grows to near-Jupiter mass and carves out a deep gap in the disk. We note that in nearly inviscid disks ($\alpha < 10^{-4}$), even deeper gaps may be opened and 
local hydrodynamic flows may become more dominant players \citep[e.g.,][]{Ginzburg18_gap}.

\begin{deluxetable*}{rcccccc}
\tablecaption{Cumulative Distribution Function of $M_{\rm gas}/M_{\rm core}$ Inferred from Observations \label{tab:cdf}}
\tablecolumns{7}
\tablewidth{0pt}
\tablehead{
\colhead{} & \colhead{$R < 2 R_\oplus$} & \colhead{Typical Sub-Neptunes} & \colhead{$R < 4 R_\oplus$} & \colhead{Typical Sub-Saturns} & \colhead{$R < 8 R_\oplus$} & \colhead{$R < 24R_\oplus$} \\
\colhead{$M_{\rm gas}/M_{\rm core}$} & \colhead{$<$0.0001} & \colhead{$<$0.01} & \colhead{$<$0.1} & \colhead{$<$0.28} & \colhead{$<$1.0} & \colhead{Jupiter\tablenotemark{a}}
}
\startdata
Case 1 & -- & $0.42^{+0.03}_{-0.03}$ & $0.84^{+0.07}_{-0.06}$ & $0.88^{+0.07}_{-0.06}$ & $0.92^{+0.07}_{-0.06}$ & 1.00 \\
Case 2 & $0.23^{+0.03}_{-0.02}$ & $0.53^{+0.04}_{-0.04}$ & $0.84^{+0.07}_{-0.06}$ & $0.88^{+0.07}_{-0.06}$ & $0.92^{+0.07}_{-0.06}$ & 1.00
\enddata
\tablenotetext{a}{When comparing to model CDFs, we truncate the model $M_{\rm gas}/M_{\rm core}$ when the total planetary mass reaches 10 Jupiter masses.}
\tablecomments{In Case 1, super-Earths are gas-stripped cores of sub-Neptunes. In Case 2, super-Earths are primordially rocky (defined as $M_{\rm gas}/M_{\rm core} < 10^{-4}$ based on the calculations by \citet{Lopez14}.}
\end{deluxetable*}

For the sub-thermal case ($R_{\rm Hill} < H$), planets are expected to be well embedded in their natal disks. Their envelopes are bound within the Bondi radius ($R_{\rm Bondi} < R_{\rm Hill}$) so that the local sound speed dominates the shear velocity. The 
maximum gas accretion rate is expected to be set by classical Bondi accretion, $\dot{M} \sim 4\pi \rho_{\rm neb} R_{\rm Bondi}^2 c_{\rm s}$. We do not consider this case because within $\sim1$ AU, even a few Earth mass cores are in the super-thermal regime. Using the temperature profile of 1000\,K\,(a/0.1\,AU)$^{-3/7}$, we find the thermal mass
\begin{equation}
    M_{\rm thermal} \sim 8\,M_\oplus \left(\frac{M_\star}{M_\odot}\right) \left(\frac{a}{0.1\,{\rm AU}}\right)^{6/7},
\end{equation}
well below the mass of cores for which accretion by local hydrodynamic flows matter (see Figure \ref{fig:Mdot_comp}). For smaller cores, their gas cooling rates are so low that they cannot reach the rate of Bondi accretion within the disk gas dissipation timescale.

\section{Distribution of \\ Gas-to-Core Mass Ratio}
\label{sec:distrb}

\subsection{Inferred from the Observed Occurrence Rates}
\label{ssec:distrb_obs}

For planets with most of their mass locked in cores, their radii are predominantly determined by their envelope mass fractions \citep{Lopez14}. For example, sub-Neptunes (2--20 $M_\oplus$, 1--4$R_\oplus$) have gas-to-core mass ratios (GCR) $\lesssim$0.1 while sub-Saturns (4--8$R_\oplus$) have GCR $\sim$0.1--1.0. The cumulative distribution function (CDF) of planet occurrence rates as a function of their radii can be thought of as the distribution of planet occurrence rates as a function of their GCR. From Figure 7 of \citet{Petigura18}, the occurrence rates of super-Earths (1--1.7$R_\oplus$), sub-Neptunes (1.7--4$R_\oplus$), sub-Saturns (4--8$R_\oplus$), and Jupiters (8--24$R_\oplus$) within $\sim$10--300 days are $17.07^{+1.85}_{-1.51}$\%, $46.44^{+3.12}_{-2.90}$\%, $5.60^{+1.45}_{-1.04}$\%, and $6.22^{+1.61}_{-1.25}$\%, respectively. 
We caution that the rates for super-Earths are a lower limit, as the sensitivity to these small planets drops beyond $\sim$75 days. Due to the small number of detected gas giants, \citet{Petigura18} only report an upper limit at $\sim$13 and $\sim$75 days, so the rates we report are also an upper limit. The occurrence rate of {\it any} planet around FGK stars within $\sim$10--300 days is then $75.33^{+4.22}_{-3.32}$\%. Since we are only interested in the relative population of planets of different sizes, we divide all occurrence rates by the total planetary occurrence rate 75.33\%. To ensure that we probe as much as possible the primordial population of planets, we do not consider planets within orbital periods of $\sim$10 days whose radii are likely altered by photoevaporation after formation \citep[see, e.g.,][their Figure 8]{Owen13}.

From \citet{Wolfgang15}, we take the median and the maximum GCR of sub-Neptunes as 0.01 and 0.1, respectively. From \citet{Petigura17}, we take the median and the maximum GCR of sub-Saturns as 0.28 and 1.0, respectively. Based on these estimates, we summarize in Table \ref{tab:cdf} the inferred cumulative distribution function of GCR for two different cases. If super-Earths are gas-stripped cores of sub-Neptunes (case 1), then there should be $(17.07\% + 46.44\%) / 2 = 31.76\%$ of planets with $M_{\rm gas}/M_{\rm core} < 0.01$. If, on the other hand, super-Earths were born as bare rocks (case 2), then the CDF at $M_{\rm gas}/M_{\rm core} < 0.01$ is $17.07\% + 46.44\%/2 = 40.29$.

\begin{figure*}
\centering
\includegraphics[width=\textwidth]{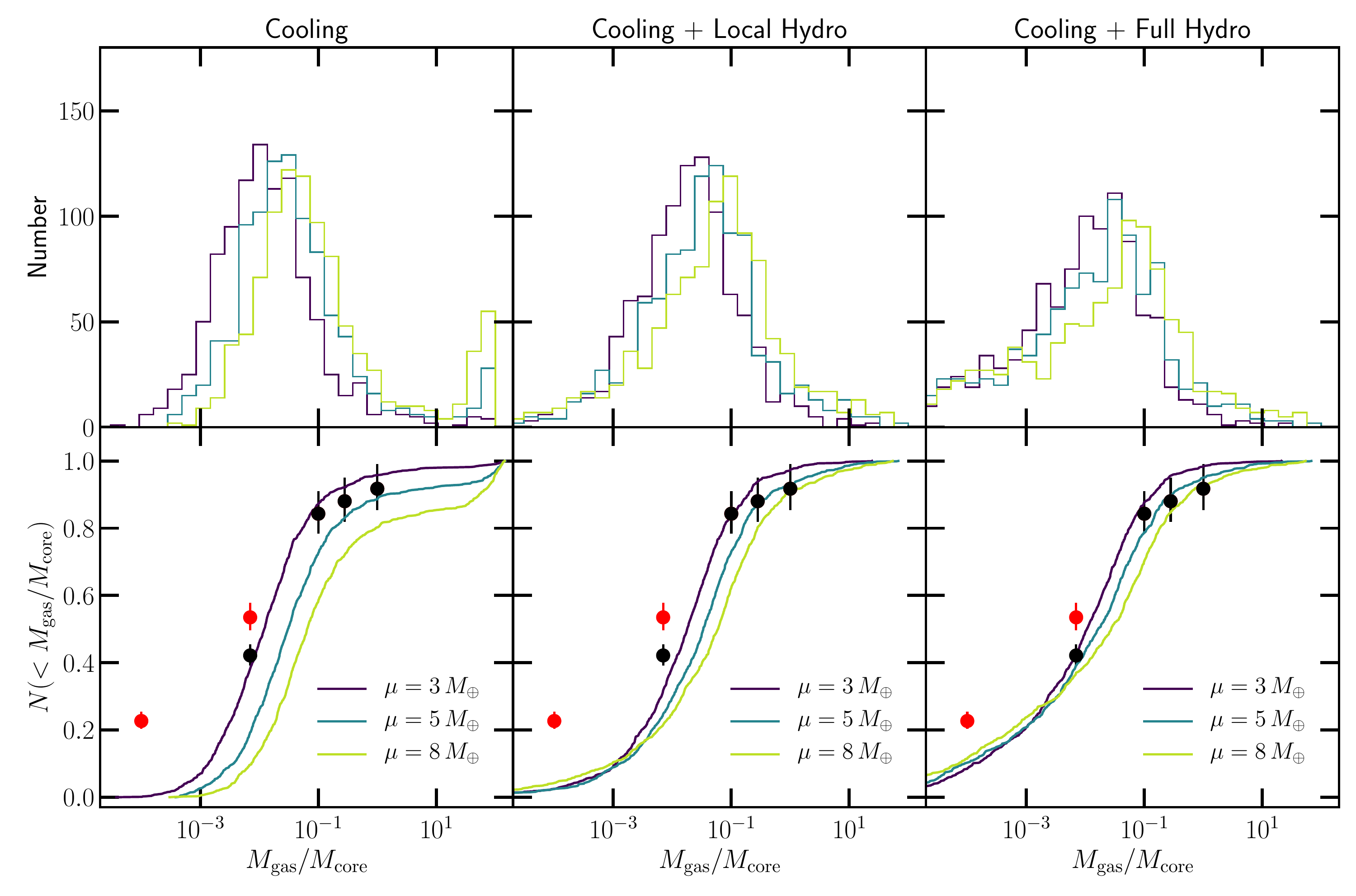}
\caption{Effect of changing median core mass on the distribution of gas-to-core mass ratios (GCR). We show both the differential (top row) and the cumulative (bottom row) distribution function. The width of the distribution is fixed to $\sigma=1$ (corresponding to 0.43 dex). All calculations are computed at 0.1 AU with $\alpha=10^{-3}$. A more massive median core mass shifts the peak GCR to a larger value. Assuming accretion to be dominated by gas cooling at all times (left column), a larger median core mass leads to a stronger secondary peak at high $M_{\rm gas}/M_{\rm core}$ corresponding to gas giants. Correcting for hydrodynamic flows (middle column) erases this secondary peak and broadens the distribution toward lower $M_{\rm gas}/M_{\rm core}$. Adding an additional correction for the global disk accretion (right column) makes the final distribution of $M_{\rm gas}/M_{\rm core}$ more bottom-heavy. Observed planet occurrence rates assuming all rocky super-Earths to be initially gas-laden sub-Neptunes (black circles) are best explained with the median core mass $\sim$3--5$M_\oplus$. If the rocky super-Earths are born rocky, we need a broader core-mass distribution.}
\label{fig:gcrdistrb_Mmed}
\end{figure*}

\begin{figure*}
\centering
\includegraphics[width=\textwidth]{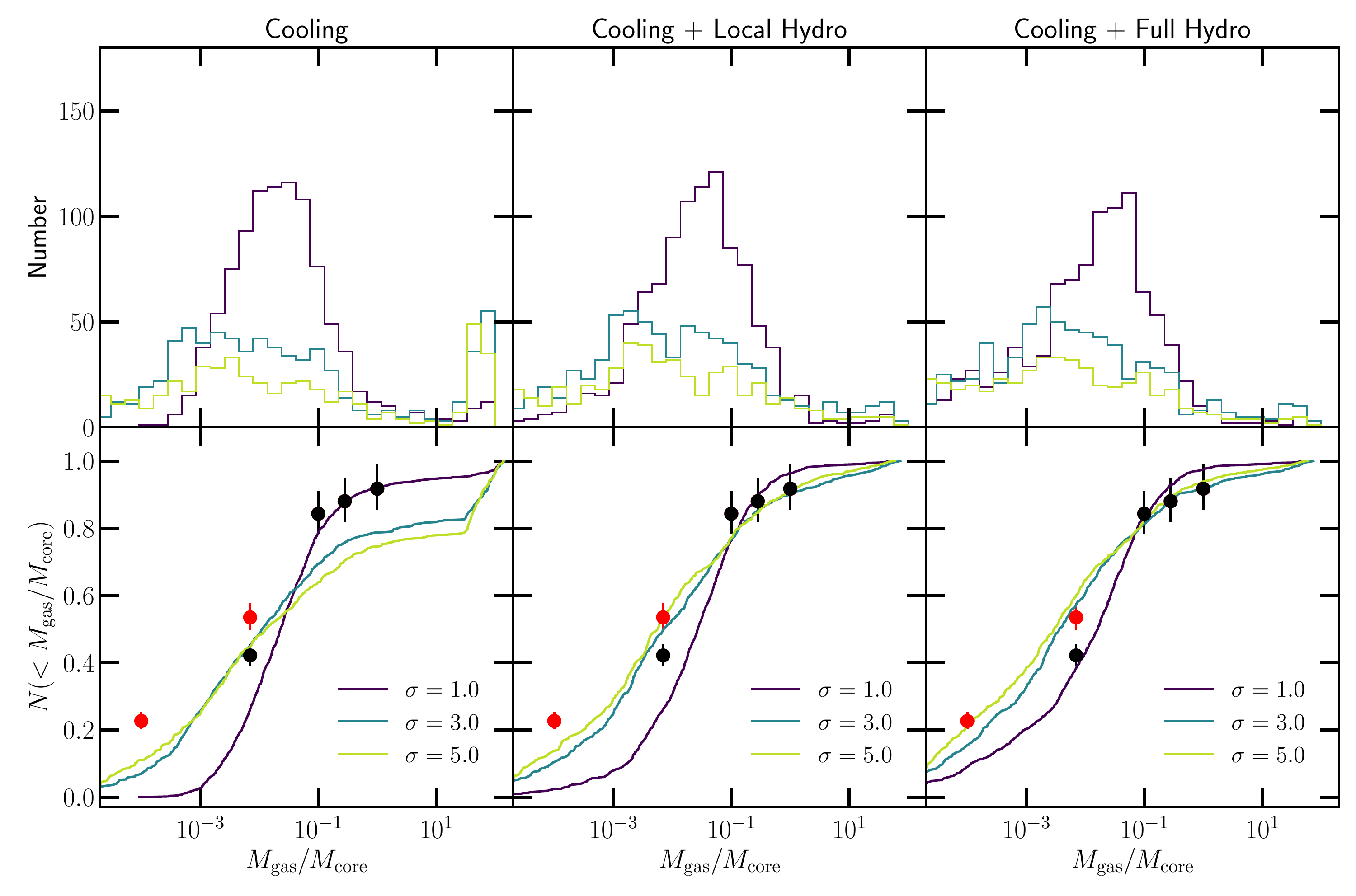}
\caption{Effect of changing the width of the core-mass distribution on the final occurrence rate of planets with different gas-to-core mass ratios $M_{\rm gas}/M_{\rm core}$. The differential and cumulative distribution functions are shown in the top and the bottom rows, respectively. The median core mass is fixed at 4$M_\oplus$. All envelope masses are computed at 0.1 AU with $\alpha=10^{-3}$. All plotting and labeling conventions mirror that of Figure \ref{fig:gcrdistrb_Mmed}. Broader distributions of core masses lead to flatter distributions of $M_{\rm gas}/M_{\rm core}$. Like Figure \ref{fig:gcrdistrb_Mmed}, we see the disappearance of the secondary peak at high $M_{\rm gas}/M_{\rm core}$ when local hydrodynamic flows are taken into account.}
\label{fig:gcrdistrb_Msig}
\end{figure*}

\subsection{Model Distribution}
\label{ssec:distrb_model}

Figures \ref{fig:gcr_comp} and \ref{fig:Mdot_comp} both demonstrate how the final gas-to-core mass ratio (GCR) is determined most sensitively by the initial core mass. We draw core masses from a lognormal distribution:
\begin{equation}
    f(M_{\rm core}) = \frac{1}{\sigma M_{\rm core} \sqrt{2\pi}} \exp{\left[-\frac{\{\ln(M_{\rm core})-\ln\mu\}^2}{2\sigma^2}\right]}
    \label{eq:fcore}
\end{equation}
where $M_{\rm core}$ is measured in $M_\oplus$, with $\mu$ as the median and the $\sigma$ as the standard deviation. We take the minimum and the maximum core masses to be 0.1$M_\oplus$ and 100$M_\oplus$, respectively.\footnote{The upper limit of 100$M_\oplus$ is motivated by the maximum possible mass of the core that may be assembled before triggering runaway gas accretion. The core assembly timescale (see equation \ref{eq:t_coag_pla}) roughly matches the gas runaway timescale for $M_{\rm core} \sim 100\,M_\oplus$.}
We draw the time over which 
cores accrete gas $t_{\rm acc}$ uniformly in linear time between 0 and 12 Myr. This upper limit is the time at which $\Sigma_{\rm neb}$ depletes by 8 orders of magnitude from the value at time zero---when the nebular gas is so tenuous that the rate of growth by cooling becomes prohibitively small \citep{Lee18}.

For each pair of $M_{\rm core}$ and $t_{\rm acc}$, we compute the final envelope mass by numerically integrating 
the gas accretion rate by cooling, hydrodynamic flows (equation \ref{eq:Mdot_hydro_norm}), or the global gas accretion rate (equation \ref{eq:Mdisk_dot}),
whichever is the smallest at each timestep of integration. Our approach closely mirrors that of \citet{Tanigawa16}.
For accretion by cooling, we multiply equation \ref{eq:gcr_exp} by $M_{\rm core}$ and take its time derivative:
\begin{equation}
    \dot{M}_{\rm cool} = M_{\rm gas}\times \left[\frac{0.4}{t} + \frac{1}{2.2\,{\rm Myrs}}\left(\frac{M_{\rm core}}{20\,M_\oplus}\right)^{4.2}\right].
    \label{eq:Mdot_cool}
\end{equation}
We do not take the time derivative of $\Sigma_{\rm neb}$ because equation \ref{eq:gcr_exp} is derived assuming static outer nebula. All our calculations are performed at 0.1 AU. Our result is insensitive to orbital distances because both $\dot{M}_{\rm cool}$ and $\dot{M}_{\rm disk}$ are spatially constant, and $\dot{M}_{\rm hydro}$ varies only weakly with distance ($\dot{M}_{\rm hydro} \propto a^{2/7}$ for massive planets that create deep gaps).\footnote{Gas accretion by cooling is spatially constant for dusty envelopes. For dust-free envelopes---defined as those whose dust grains do not contribute to the overall atmospheric opacity---$\dot{M}_{\rm cool}$ rises farther away from the star where the disk is cold and the vibrational degrees of freedom in molecular species freeze out \citep{Inamdar15,Lee15,Piso15}. The envelope becomes more transparent and so cools faster.} For low-mass planets that carve out a shallow gap (or no gap at all), although $\dot{M}_{\rm hydro} \propto a^{-8/7}$, it is still orders of magnitude higher than $\dot{M}_{\rm cool}$ so that their gas accretion is cooling-limited.

All envelopes stop growing once they develop isothermal profiles with their temperatures set to that of the outer nebula---at this point, the entire envelope has reached thermal equilibrium with the outer nebula and cools no more. For sub-Earth mass cores, this maximally cooled state is reached at GCRs well below that computed using equation \ref{eq:gcr_exp}. For all model planets with $M_{\rm core} \leq 1\,M_\oplus$, we cap their envelope masses to that dictated by their isothermal profiles. We also impose the absolute maximum total mass of all planets to be 10 Jupiter masses, where there is an evidence of a ``desert'' in the mass function of substellar objects \citep[e.g.,][]{Schlaufman18}, separating massive gas giants from low-mass brown dwarfs.

Figures \ref{fig:gcrdistrb_Mmed} and \ref{fig:gcrdistrb_Msig} illustrate the effects of varying core-mass distributions. Increasing the median core mass shifts the primary peak of the GCR distribution to a larger value. This peak is set by the amount of gas accreted by the median core mass over the median accretion time $\sim$6 Myr; a top-heavy core-mass distribution will lead to a top-heavy GCR distribution.
If gas accretion is cooling-limited for all cores at all times, we find a secondary peak at GCR $\gtrsim$10 because the runaway accretion proceeds uninhibited. Accounting for the local hydrodynamic flows halts the runaway and mutes the secondary peak, bringing the model GCR distribution to a closer resemblance to the observed distribution of planetary radii, which shows no obvious peak at radii beyond 4$R_\oplus$ within $\sim$100 days \citep[see, e.g.,][their Figure 5]{Fulton18}. We find that accounting for the global disk accretion in addition to local hydrodynamic flows does not alter the high-end tail of the GCR distribution. Instead, the entire GCR distribution becomes more bottom-heavy. This extra gas-poor population comes from cores that assembled late during the rapid disk gas dispersal. The disk accretion rate falls below $\sim$10$^{-10}\,M_\odot\,{\rm yrs}^{-1}$ and dictates how much gas cores can accrue. 
All cores attain the same amount of gas mass over a given timescale in this phase so that the GCR is particularly small for the more massive cores. 
Our result is robust against different choices of the initial rate of disk accretion as long as the background disk undergoes a stage of late-time rapid gas dispersal. 

Since the mass of the core is the strongest determinant of how rapidly gas can be accreted, the width of the underlying core-mass distribution is directly commensurate with the width of the GCR distribution.
Overall, the sharp rise in the inferred cumulative distribution of GCR points to a sharp peak in the differential distribution at GCR$\sim$0.01. The strength and the sharpness of the peak suggest that the distribution of core masses cannot be too broad (see Figure \ref{fig:gcrdistrb_Msig}). 
If we insist the rocky planets (1--1.7$R_\oplus$) were born rocky, a very broad (almost uniform) distribution of core mass is required to explain the observed occurrence rates. Figure \ref{fig:gcrdistrb_Msig} shows that for $\mu=4M_\oplus$, $\sigma$ needs to be at least $\sim$3 (equivalent to 1.3 dex) and closer to $\sim$5 (equivalent to 2 dex). We find such broad distributions unlikely, as the radial velocity follow-up of {\it Kepler} planets reports a sharp drop in planet masses beyond $\sim$10$M_\oplus$ \citep{Marcy14}. It is more probable that a separate population of low-mass, bare rocky objects were created in the gas-free era, after all the nebular gas was exhausted, analogous to the formation of terrestrial planets in the solar system \citep[see also][]{Owen18}. It is also possible that various envelope-loss mechanisms such as giant impact \citep{Inamdar16}, photoevaporation by high-energy photons from the star \citep{Lopez12,Owen13}, and/or envelope-powered or core-powered mass loss \citep{Owen16,Ginzburg18} created these rocky planets beyond $\sim$10 days (with the caveat that the latter two mechanisms lose their potency at longer orbital periods). We discuss in more detail the possible origin of these rocky planets in Section \ref{sec:summary}.

\section{Best-fit Core Mass Distribution}
\label{sec:best-fit}

We now search for the best-fit core mass distribution. The model distribution of gas-to-core mass ratio (GCR) is computed on a grid of 100 $\mu$ $\times$ 100 $\sigma$, evenly and linearly spaced between 1 and 30$M_\oplus$ and 0.1 to 2.5, respectively. We initially draw 10,000 cores from each distribution and remove any core with mass below 0.1$M_\oplus$ or above 100$M_\oplus$. For each core, we assign accretion timescales randomly and uniformly drawn in linear time from 0 to 12 Myr. The ``best-fit'' distribution is defined as the one that maximizes the likelihood function:
\begin{equation}
    \log\mathcal{L} = -\frac{1}{2}\Sigma_i\left[\log(2\pi\sigma_{\rm obs,i}^2)+\frac{(C_{\rm model,i}-C_{\rm obs,i})^2}{\sigma_{\rm obs,i}^2}\right]
    \label{eq:likelihood}
\end{equation}
where $C_{\rm model}$ is the model CDF of GCRs, $C_{\rm obs}$ is the inferred CDF from the observed planet occurrence rates, $\sigma_{\rm obs,i}$ are their errors, and i iterates over GCRs of 0.01, 0.1, 0.28, and 1.0 (see Table \ref{tab:cdf}, case 1). 

\begin{figure}
    \centering
    \includegraphics[width=\columnwidth]{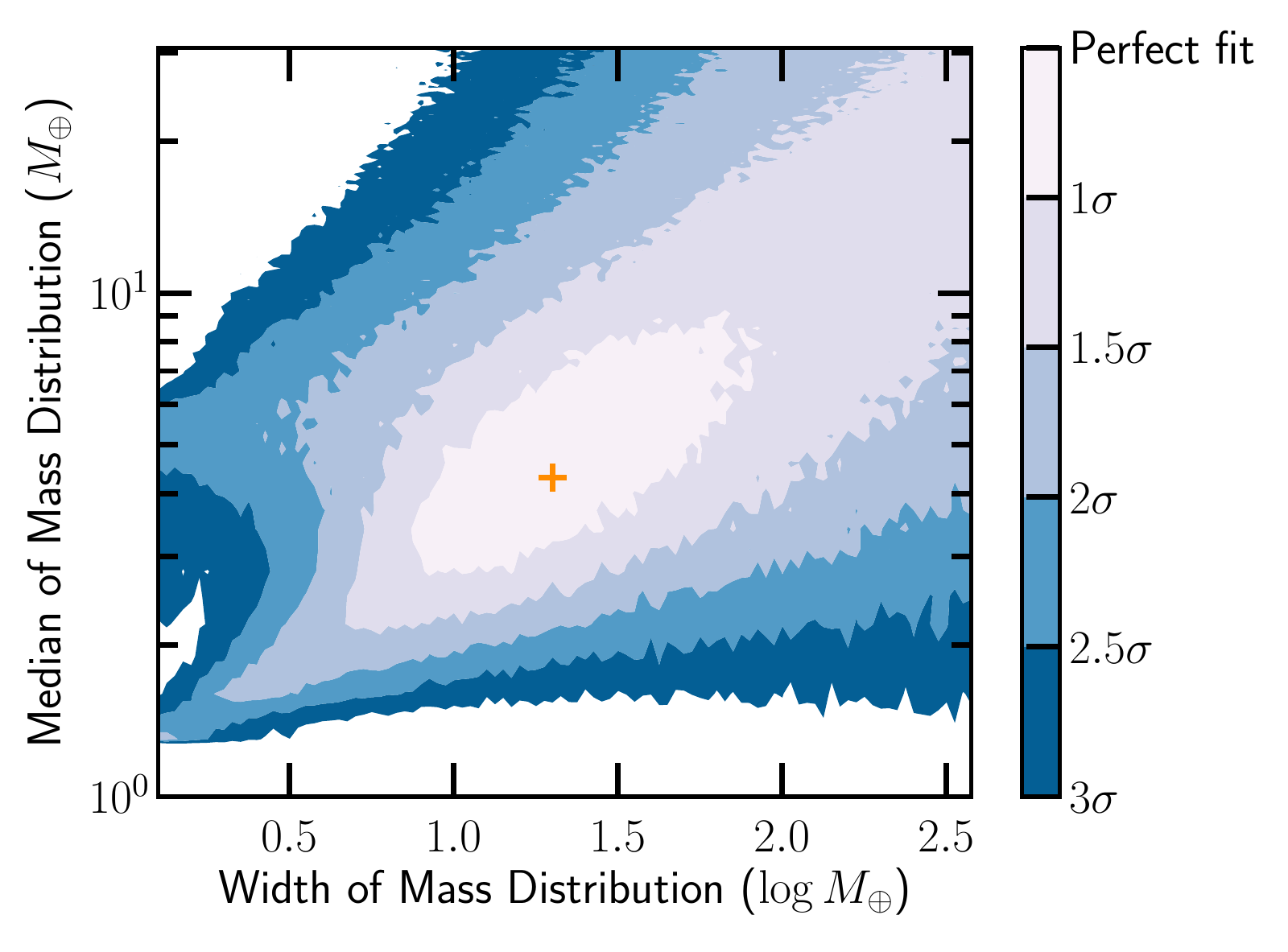}
    \caption{Likelihood of the model cumulative distribution of gas-to-core mass ratios matching the observations (equation \ref{eq:likelihood}).
    Accretion times are drawn uniformly in linear space from 0 to 12 Myr. The core-mass distribution that maximizes the likelihood function is marked with an orange cross: median of 4.30$M_\oplus$ and the standard deviation of 1.30 (equivalent to 0.56 dex). There is a positive correlation between $\mu$ and $\sigma$ because for high medians, the core-mass distribution needs to be broad enough to encompass smaller cores that nucleate sub-Neptunes.}
    \label{fig:Mdistrb_fit}
\end{figure}

We find the best-fit core mass distribution to be described by a lognormal distribution with the median of 4.30\,$M_\oplus$ and the standard deviation of 1.30 (equivalent to 0.56 dex). Figure \ref{fig:Mdistrb_fit} shows a positive correlation between the best-fit $\mu$ and $\sigma$. The maximum likelihood map reflects how gas accretion by cooling predicts typical sub-Neptune cores to be less than 10$M_\oplus$, consistent with the radial velocity measurements \citep{Marcy14,Weiss14}. When the median core mass is large, the distribution needs to be broad enough to have a sufficient number of small cores to reproduce the ubiquity of sub-Neptunes.

This median core mass can be easily understood by the requirement that the median GCR has to be $\sim$0.02 to fit the observations. A linear sampling of accretion time from 0 to 12 Myr corresponds to the median time of 6 Myr. From equation \ref{eq:gcr}, we find that $M_{\rm core} \sim 4\,M_\oplus$ achieves $M_{\rm gas}/M_{\rm core} \sim 0.03$ within 6 Myr. 

\begin{figure*}
    \centering
    \includegraphics[width=\textwidth]{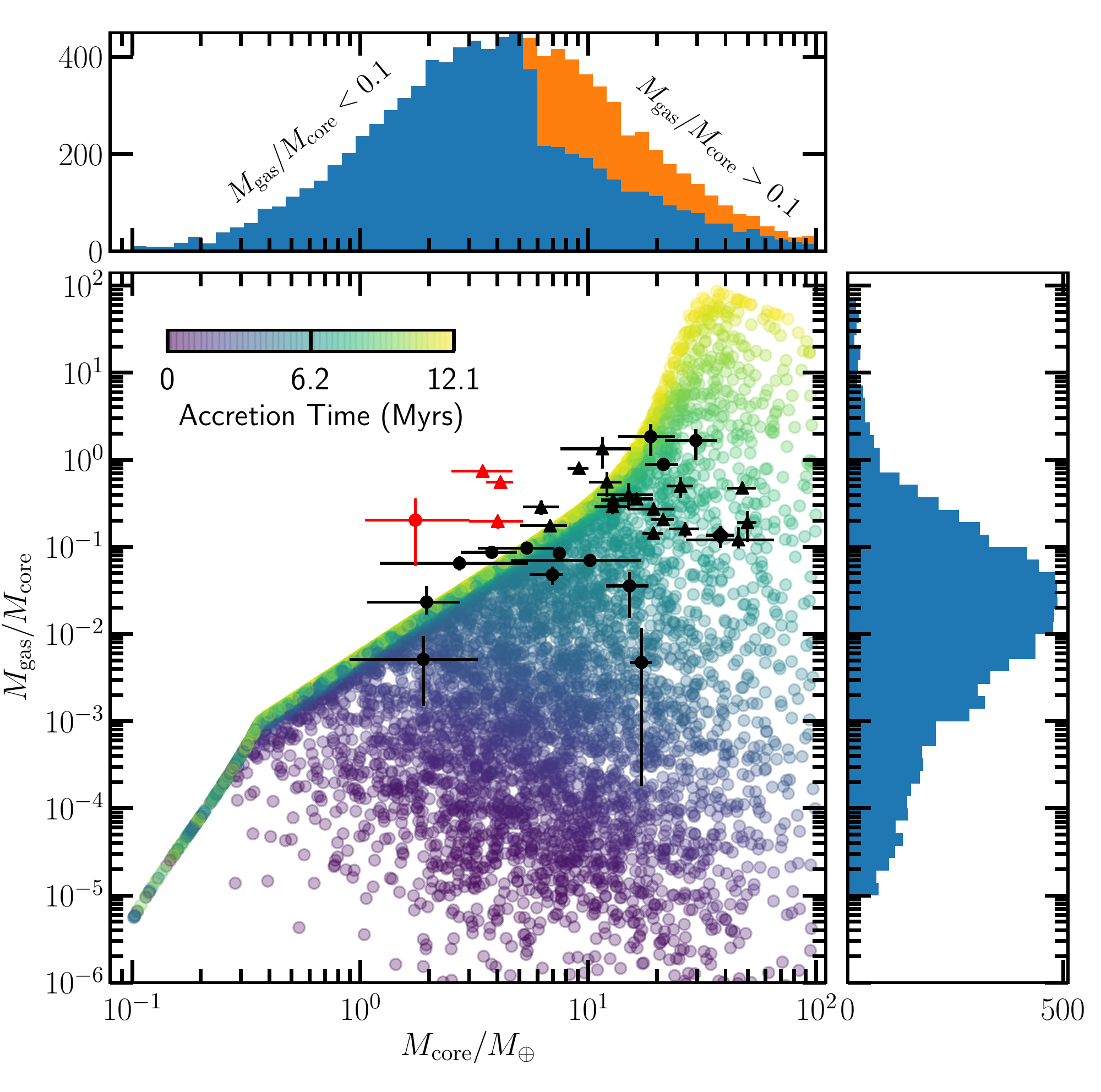}
    \caption{Final gas-to-core mass ratio ${\rm GCR} \equiv M_{\rm gas}/M_{\rm core}$ given the core mass $M_{\rm core}$ and the accretion time (color-coded) for the best-fit model ensemble of planetary core masses ($\mu=4.3M_\oplus$, $\sigma=1.3$). We confirm the typical $M_{\rm gas}/M_{\rm core}$ $\sim$ 0.01 (the right histogram). We see four distinct regimes of maximum GCR. Cores with $M_{\rm core} \lesssim 0.4M_\oplus$ cannot accrete beyond the isothermal maximally cooled state. For $0.4 \lesssim M_{\rm core}/M_\oplus \lesssim 10$, envelope cooling proceeds until the dispersal of the nebular gas. Beyond $10M_\oplus$, planets are able to enter the runaway regime but their growth is ultimately stymied by local hydrodynamic flows. Cores larger than $\sim$40$M_\oplus$ are so massive that they carve out a deep gap, and the maximum $M_{\rm gas}$ they can attain drops with core mass. The median core mass of sub-Saturns and Jovians ($M_{\rm gas}/M_{\rm core} > 0.1$ in the upper panel) roughly corresponds to the mass at which the limiting mechanism of envelope growth switches from gas cooling to hydrodynamic flows. Real-life exoplanets are plotted in circle \citep{Lopez14} and triangle \citep{Petigura17} points. We also plot K2-55b \citep{Dressing18} in diamond. Apart from super-puffs (marked in red; Kepler-51b, Kepler-223e, Kepler-87c, Kepler-79d), data points fall within the expected range of $M_{\rm gas}/M_{\rm core}$ for given $M_{\rm core}$. These super-puffs require a special condition that not only do they have to build their envelopes beyond $\sim$1 au but also that their accretion needs to be dust-free \citep{Lee16}. There are a few data points that lie slightly above the expected maximum $M_{\rm gas}/M_{\rm core}$; these planets have similarly low bulk densities as the known super-puffs ($\rho_{\rm bulk} \lesssim 0.5\,{\rm g\,cm^{-3}}$) and so may share the same formation history.}
    \label{fig:gcr_Mcore_time}
\end{figure*}

\section{The Final Outcome}
\label{sec:outcome}

In principle, all cores that are more massive than the Earth can become gas giants if they build their envelopes in an infinite reservoir of nebular gas. In reality, protoplanetary disks do not retain their gas for an indefinite amount of time \citep{Mamajek09,Alexander14}. The fate of a planet is sealed at birth by the properties of its core: how massive it is and when it was assembled with respect to the disk gas dissipation timescale.

Figure \ref{fig:gcr_Mcore_time} illustrates five different regimes of the maximum envelope mass fraction a core can attain. Gas-rich sub-Saturns/Jupiters ($M_{\rm gas}/M_{\rm core} > 1.0$) are nucleated only by cores that are more massive than $\sim$10$M_\oplus$.\footnote{There is a small population of sub-Saturns with the inferred core masses as small as $\sim$2$M_\oplus$. These super-puffs must have initially formed as dust-free worlds outside $\sim$1 AU \citep{Lee16}. Post-formation dust enrichment of their envelopes can potentially explain their present-day flat transmission spectra \citep[see, e.g.,][and references therein]{Wang19}.} These are massive enough that they can trigger the runaway gas accretion well within the disk lifetime (defined as the 
so-called
``critical'' core mass; \citealt{Pollack96,Rafikov06}).
Runaway is launched but never prolongs. 
The global disk accretion during the phase of late-stage rapid dispersal ultimately limits the growth of planets, generating sub-Saturns as failed Jupiters (see also Figure \ref{fig:Mdot_comp}).

Above $\sim$40$M_\oplus$, the maximum GCR falls with increasing core mass. These cores are so massive that they can carve out a deep gap: the local depletion factor is at least $\sim$200 at 0.1 AU, with a disk temperature of 1000 K, and will be even more substantial for larger cores (see equation \ref{eq:K}). Cores that weigh $\sim$40--80$M_\oplus$ trigger the runaway but the reduction in the density of the inflow gas limits the maximum GCR for these gargantuan cores. Cores that are heavier than 80$M_\oplus$ can only build their envelopes at a rate set by the global disk accretion.

On the opposite spectrum below $\sim$10$M_\oplus$, the maximum GCR is set by how much a given core can accrete by cooling within the full disk lifetime of 12 Myr. Down to an Earth mass core, the maximum GCR ranges between $\sim$10$^{-3}$ and $\sim$10$^{-1}$ so these cores are guaranteed to become gas-poor sub-Neptunes. Those that assemble earlier in the disk lifetime will gather more gas and become puffier, while those that assemble later will remain almost bare rocks. Cores below $\sim$0.4$M_\oplus$ are so tiny that their maximum possible GCR (described by maximally cooled isothermal envelope) is well below $\sim$10$^{-3}$. 

Like the model, we find that the observed exoplanets with more massive cores tend to have larger $M_{\rm gas}/M_{\rm core}$ (see Figure 6; data from \citet{Lopez14} and \citet{Petigura17}). We limit the sample to those with orbital periods longer than 10 days so as to minimize the effect of envelope mass loss \citep[e.g.,][]{Owen16, Ginzburg18} and radius inflation \citep[e.g.,][]{Weiss13,Thorngren18}. We find a few planets that feature at least an order of magnitude larger than the expected maximum $M_{\rm gas}/M_{\rm core}$. These are known super-puffs (Kepler-51b, \citealt{Steffen13}, \citealt{Masuda14}; Kepler-223e, \citealt{Mills16}; Kepler-87c, \citealt{Ofir14}; Kepler-79d, \citealt{Jontof-hutter14}) whose extreme low densities require a special formation condition: dust-free gas accretion beyond $\sim$1 au \citep{Lee16}.

\section{Summary and Discussion}
\label{sec:summary}

Sub-Saturns (4--8$R_\oplus$) are inferred to have envelope mass fractions $M_{\rm gas}/M_{\rm core} \sim 0.1$--1.0, just at the edge of runaway gas accretion. Theories of core accretion expect a significant excess of gas giants relative to sub-Saturns. Yet, the observed occurrence rates of these two classes of gas-rich planets are comparable \citep[e.g.,][]{Dong13,Petigura18}. We have shown that this similarity between sub-Saturns and gas giants can be robustly explained if local hydrodynamic flows and the global disk gas accretion are taken into account, both of which significantly stall the rate of gas delivery to the planet, effectively shutting off the runaway.

How rapidly planets can build their envelopes is most sensitively determined by their core masses. The ubiquity of sub-Neptunes and the scarcity of gas-rich planets inform a distribution of underlying core masses that is peaked toward $<10M_\oplus$. We report a lognormal distribution with a median of 4.3$M_\oplus$ and the standard deviation of 1.30 in $\log M_{\rm core}$ (equivalent to 0.56 dex) that can self-consistently explain the occurrence rates of sub-Neptunes, sub-Saturns, and Jupiters. These values are similar to that inferred from the photoevaporation model \citep{Owen17} fitted to the observed radius gap \citep{Fulton17} as well as radial velocity follow-up of {\it Kepler} sub-Neptunes \citep[e.g.,][]{Marcy14}.

We have identified four different regimes of maximum envelope mass fraction based on four different ranges of core masses. The process that determines this maximum fraction switches from gas thermodynamics to hydrodynamics at $M_{\rm core}\sim$10$\,M_\oplus$, breeding gas-rich sub-Saturns and Jupiters. Beyond $M_{\rm core} \sim 40\,M_\oplus$, a further increase in core masses reverses the growth of envelopes as they carve out deep gaps, reducing the local nebular density by orders of magnitude. Below $\sim$10$M_\oplus$, accretion by cooling in a gas disk that is dispersing over timescales of $\sim$10 Myr guarantees the formation of sub-Neptunes and super-Earths. Tiny cores below $\sim$0.4$M_\oplus$ can only build at maximum $M_{\rm gas}/M_{\rm core} \lesssim 10^{-3}$, as dictated by their isothermal, maximally cooled atmospheres.

Below, we discuss how our theory might bear on other properties of {\it Kepler} planets and identify avenues for improvements.

\subsection{The Diversity of Planets in Metal-rich Systems}
\label{ssec:ssat-snep}

Figure \ref{fig:gcr_Mcore_time} shows that for typical disks that deplete their gas rapidly after $\sim$1 Myr, only the cores that are more massive than $\sim$5$M_\oplus$ and typically $\sim$10$M_\oplus$ can become sub-Saturns and Jupiters (see also Section \ref{sec:outcome}). On the other hand, sub-Neptunes can appear from a wide range of core masses since massive cores that assemble late do not have enough time to trigger runaway accretion. Figure \ref{fig:gcr_Mcore_time} illustrates that the range of core masses that can nucleate sub-Neptunes is larger by at least an order of magnitude compared to the range of core masses that can nucleate sub-Saturns and gas giants.

The wide range of possible origins for sub-Neptunes may be the reason why their occurrence rates do not correlate strongly with the host star metallicity. Numerous studies report a strong correlation between the stellar metallicity and the occurrence rate of Jupiters \citep[e.g.,][]{Fischer05} and sub-Saturns \citep[e.g.,][]{Petigura18} but considerably weaker correlation for smaller sub-Neptunes \citep[e.g.,][]{Buchhave14,Wang15}. It is likely that metal-rich stars harbored solid-heavy disks that carry the potential of creating more massive cores. Because gas-rich sub-Saturns and Jupiters require massive cores, they are more likely to be found in solid-heavy disks and therefore around more metal-rich stars. These same cores, if they assemble late, cannot accrete enough gas and end up gas-poor. In other words, massive cores have the potential to become either gas-rich or gas-poor, whereas small cores can only become gas-poor. An equivalent interpretation is that metal-rich systems harbor a wider variety of planets. Follow-up radial velocity surveys are required to verify whether the core-mass distribution of sub-Neptunes ($<4\,R_\oplus$) beyond $\sim$10 days is indeed wider than those of sub-Saturns.

\subsection{Core Assembly Time and Intra-System Uniformity}
\label{ssec:core_assembly}

We have assumed cores of all masses can appear at any time. The fact that there are at least factors of $\sim$3 scatter in the mass-radius relationship of {\it Kepler} super-Earths/mini-Neptunes \citep{Wolfgang15} suggests that for a given core mass, there must be some variation in their assembly times. Although post-formation---i.e., after the nebular gas completely disperses---giant impact can also produce such variation in the mass-radius relationship \citep[e.g.,][]{Inamdar16}, collisions in the absence of ambient gas can potentially result in planet pairs of dissimilar densities. Most {\it Kepler} multi-planetary systems feature planets with similar radii and masses \citep{Millholland17,Weiss18}, suggesting any system-to-system variations we observe are signatures of varying formation environments, rather than processes that occur after formation.

The time at which cores appear depend on their assembly process.
Planetary cores can be built via minor mergers (pebble and planetesimal accretion) and/or major mergers (giant impact). The former requires gas-rich environment, while the latter is more likely to proceed in gas-poor---but not gas-empty---nebula. The main difference between pebble and planetesimal accretion is the size of the solid particles that are being accreted by the seed core. ``Pebbles'' refer to particles with Stokes numbers (the number of local orbital time it takes for particles to reach their terminal velocities) near unity. Gas aerodynamic drag can efficiently damp away pebbles' random velocities to increase the accretion cross section of the seed cores and so boost the rate of core growth \citep{Ormel10,Lambrechts12}. Planetesimals are large enough to be decoupled from the gas flow and their dynamics are largely unaffected by the gas drag. 

Within $\sim$1 AU where we are interested in, the dynamical clock runs so fast that both pebble accretion and planetesimal accretion can proceed within $\sim$ $10^4$ years. In the most pessimistic scenario, the minimum accretion cross section is given by the core's geometric cross section, and the time it takes to build up to $\sim$4$M_\oplus$ is
\begin{align}
	t_{\rm pla} &\sim \frac{M_{\rm core}}{\dot{M}} \sim \frac{M_{\rm core}}{\Sigma_{\rm s}R_{\rm core}^2\Omega} \nonumber \\
	&\sim \left(\frac{a}{R_{\rm core}}\right)^2\Omega^{-1} \nonumber \\
	&\sim 10^{4}\,{\rm yrs}\left(\frac{a}{0.1\,{\rm AU}}\right)^{3.5}\left(\frac{1.5\,R_\oplus}{R_{\rm core}}\right)^{2}.
	\label{eq:t_coag_pla}
\end{align}
where we approximated the solid surface density $\Sigma_{\rm s} \sim M_{\rm core}/a^2$ and computed the core radius using $R_{\rm core} \sim R_\oplus (M_{\rm core}/M_\oplus)^{1/3}$. Using $R_{\rm core} \propto M_{\rm core}^{1/4}$ provides a similar result. The accretion timescale can shorten if there are more planetesimals in the disk (i.e., if $\Sigma_{\rm s}$ is higher).

The timescale of pebble accretion can be shorter than $t_{\rm pla}$ by orders of magnitude, but it is a self-limiting process. Once cores grow to masses large enough so that their tidal torques overcome the viscous torque of the surrounding gas, gaps can be carved out in the gas nebula. The outer edge of this gap acts as a dust trap, barricading the cores from further influx of pebbles \citep[e.g.,][]{Lambrechts14}. \citet{Bitsch18} reported this pebble isolation mass using an empirical fit to their numerical simulations:
\begin{equation}
M_{\rm peb, iso} \sim 1.4\,M_\oplus\left(\frac{H/a}{0.02}\right)^3\left[0.34\left(\frac{-3}{\log_{\rm 10}\alpha}\right)^4 + 0.66\right],
\end{equation}
where the disk aspect ratio $H/a = 0.02$ is evaluated at 0.1 AU around a solar mass star with a disk temperature of $1000\,{\rm K}(a/0.1\,{\rm AU})^{-3/7}$. 
The pebble isolation mass is just a few Earth masses within 1 AU for $\alpha = 10^{-3}$ and can drop below an Earth mass for $\alpha < 10^{-4}$ \citep{Fung18}. Another feature of the pebble isolation mass is that it does not depend on the solid mass reservoir (as long as the disk contains enough solids to create cores of pebble isolation mass). It is unlikely that the final planetary cores of inner Kepler planets are set by pebble isolation, as it is difficult to reconcile with the observed correlation between stellar metallicity and the planet occurrence rate (\citealt{Wang15,Petigura18,Owen18} but see \citealt{Wu18} for an opposing view). Furthermore, pebble isolation mass rises steeply with orbital distance ($M_{\rm peb,iso} \propto a^{6/7}$ for our choice of temperature profile), with no other stronger dependence on either the disk property or the stellar mass. This is hard to reconcile with the observed intra-system uniformity in mass and radius. Either a constant supply of larger particles whose inflows are uninhibited by perturbations in the gas disk is required or major mergers between neighboring bodies are required.

We now consider the scenario where the final planetary cores are built by giant impact when there is still some gas around. Because of the faster dynamical clock closer to the star, giant impact proceeds more rapidly there so that any initial rise in planet mass toward longer orbital periods flattens \citep[see, e.g.,][their Figure 1]{Dawson15}, potentially explaining the observed intra-system uniformity in the masses of {\it Kepler} multi-planetary systems \citep{Millholland17}.
Planetary radii are largely determined by the gas mass fraction, which in turn is largely governed by the core mass. In the theory of core accretion, intra-system uniformity in radii \citep{Weiss18} follows directly from the intra-system uniformity in masses as long as the core assembly is complete before the disk gas disperses. Giant impacts in a gas-free era should be rare.

From equating the orbital crossing timescale to the gas eccentricity damping timescale, \citet{Lee16} found that to produce present-day Kepler multi-planet systems by major mergers, the nebular gas needs to be depleted by at least four orders of magnitude with respect to the solar nebula \citep[see also][]{Kominami02}. For the disk model that we use (equation \ref{eq:Sigma_gas}), such a level of depletion is reached at $\sim$6.4 Myr. Cores that are assembled at this time would have $\sim$5.7 Myr to accrete gas and would build $M_{\rm gas}/M_{\rm core}$ $\sim$ 0.1 at best (see Figure \ref{fig:gcr_Mcore_time}). This is consistent with the argument made by \citet{Lee16} that super-Earths and mini-Neptunes can robustly emerge in late-stage, gas-poor nebula. Bare rocky planets would assemble even later when there is practically no gas left (see Figure \ref{fig:gcr_Mcore_time} for accretion time $\sim$0 Myr). Sub-Saturns and gas giants, on the other hand, would have to nucleate from massive ($>10M_\oplus$) cores that assembled early. It may be that their cores assembled by pebble accretion farther out in the disk where the pebble isolation mass is higher, then migrated in. It may also be that they are the product of rare collisions in the inner disk that occurred in the gas-rich era. Distinguishing between these two scenarios is a subject of future work.

\acknowledgements
I thank Konstantin Batygin, Eugene Chiang, Sivan Ginzburg, Brad Hansen, Phil Hopkins, Andrew Howard, Heather Knutson, Dong Lai, Erik Petigura, Jason Wang, and Yanqin Wu for helpful discussions. The anonymous referee provided a careful report that helped to improve the manuscript. E.~J.~L is supported by the Sherman Fairchild Fellowship at Caltech. This research used the Savio computational cluster resource provided by the Berkeley Research Computing program at the University of California, Berkeley (supported by the UC Berkeley Chancellor, Vice Chancellor for Research, and Chief Information Officer).

\bibliography{cmf}
\end{document}